\documentclass[pdftex,twocolumn,epjc3]{svjour3}          

\RequirePackage[T1]{fontenc}

\smartqed  

\RequirePackage{graphicx}
\RequirePackage{mathptmx}      
\RequirePackage{flushend}
\RequirePackage[numbers,sort&compress]{natbib}
\RequirePackage[colorlinks,citecolor=blue,urlcolor=blue,linkcolor=blue]{hyperref}

\newcommand{\wildcard}{$\langle*\rangle$}

\usepackage{algorithm}
\usepackage{algpseudocode}
\usepackage{amssymb}

\journalname{International Journal of Information Security}

\begin{document}

\title{Using Large Language Models for Template Detection from Security Event Logs}

\author{Risto Vaarandi\thanksref{e1,addr1}
        \and
        Hayretdin Bah\c{s}i\thanksref{e2,addr1,addr2} 
}

\thankstext[$\star$]{t1}{Corresponding author: Risto Vaarandi}
\thankstext{e1}{e-mail: risto.vaarandi@taltech.ee}
\thankstext{e2}{e-mail: hayretdin.bahsi@taltech.ee}

\institute{Centre for Digital Forensics and Cyber Security, Tallinn University of Technology, Estonia\label{addr1}
          \and
          School of Informatics, Computing and Cyber Systems, Northern Arizona University, USA\label{addr2}
}

\date{}

\maketitle

\begin{abstract}
In modern IT systems and computer networks, real-time and offline event log analysis is a crucial part of cyber security monitoring. In particular, event log analysis techniques are essential for the timely detection of cyber attacks and for assisting security experts with the analysis of past security incidents. The detection of line patterns or templates from unstructured textual event logs has been identified as an important task of event log analysis since detected templates represent event types in the event log and prepare the logs for downstream online or offline security monitoring tasks. During the last two decades, a number of template mining algorithms have been proposed. However, many proposed algorithms rely on traditional data mining techniques, and the usage of Large Language Models (LLMs) has received less attention so far. Also, most approaches that harness LLMs are supervised, and unsupervised LLM-based template mining remains an understudied area. The current paper addresses this research gap and investigates the application of LLMs for unsupervised detection of templates from unstructured security event logs.
\end{abstract}

\section{Introduction}\label{section_introduction}

\begin{figure}[htbp]
\centerline{\includegraphics[width=0.5\textwidth]{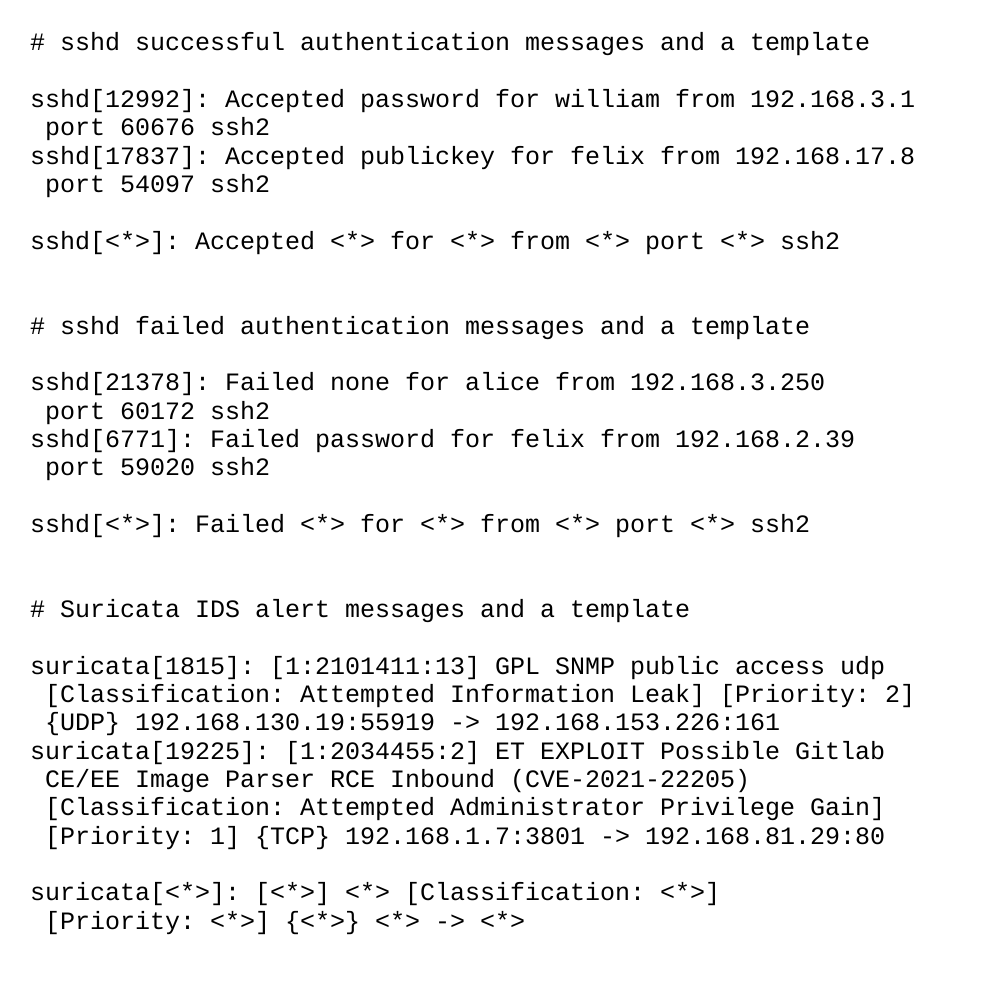}}
\caption{Example \emph{syslog} messages and their templates}
\label{example-messages}
\end{figure}

Event log analysis is an important cyber security monitoring technique. For example, modern Security Operations Center (SOC) platforms like Splunk \cite{splunk} and ElasticStack \cite{elasticstack} rely on collecting and analyzing event logs from a large number of sources, whereas dedicated event correlation tools like SEC \cite{sec} have been designed for real-time monitoring of event logs. Although during recent years several structured event log formats have been proposed, a significant part of event log messages remain unstructured or semi-structured. As an example, the widely used BSD \emph{syslog} protocol \cite{bsd-syslog} defines the priority, timestamp, hostname, and \emph{syslog} tag (program name with its process ID) fields, whereas the message text, which is the most informative part of the \emph{syslog} message, remains a free-form string. Figure \ref{example-messages} depicts some example \emph{syslog} messages without the priority, timestamp, and hostname fields (note that all \emph{syslog} messages have been wrapped in Figure \ref{example-messages} due to their length, and they appear in one line in real-life event logs).

As Figure \ref{example-messages} illustrates, some parts of textual event log messages have a constant nature, while other parts are variables (such as IP addresses, process IDs, and port numbers). Template detection (or template mining) algorithms assume that each message in the event log is represented by a single line, and template detection involves reporting line patterns (templates) to the end user, which summarizes the content of the event log file. Each reported template should represent a group of event log messages, where constant parts shared by all messages are reported as is, and variable parts of messages are replaced with wildcards (in many past research papers, \wildcard{} has been used for denoting the wildcard, and we have adopted this notation in the current paper). In addition to example \emph{syslog} messages, Figure \ref{example-messages} also depicts relevant templates for similar messages. The mining of such templates is highly useful for several purposes, such as identifying event types in event logs, writing parsers (e.g., regular expression-based parsing rules) for these event types, and preparing unstructured event log data for downstream machine learning tasks.

During the last two decades, a number of template detection algorithms have been proposed \cite{drain, logsig, iplom, spell, ael, lenma, logcluster, slct, lfa, lke, logmine, shiso, molfi, logram, logppt, divlog, lilac, llmparser, ulog}, and several comparative studies have been published for these algorithms \cite{khan, zhu, gasimov}. Despite a significant amount of research in the domain of template mining, some research gaps have remained. 
Firstly, template detection approaches that have not employed LLMs (we call them \emph{traditional algorithms} in the remainder of this paper) are unsupervised data mining-based approaches which can be tuned through hyperparameters. However, as pointed out in \cite{logppt}, selecting optimal values for hyperparameters is often not an easy task. Secondly, LLM-based template detection approaches are generally supervised and require labeled training data \cite{llmparser, divlog, lilac, logppt}. As discussed in \cite{ulog}, the creation of such training data is a laborious process, and training data sets need to be updated on the appearance of new event log message types.

Thirdly, many existing template mining approaches have various design limitations. As pointed out in recent studies \cite{gasimov, logppt}, several algorithms rely on preprocessing event log messages, which requires previous domain knowledge of message formats. Also, as discussed in \cite{gasimov, divlog}, a number of algorithms incorrectly assume that the log messages described by the same template always contain the same number of words. When that is not the case (for example, two Suricata IDS alert messages from Figure \ref{example-messages} consist of 17 and 23 words, respectively), the algorithms fail to detect a common template for such messages. 

Finally, existing studies have often used data sets which are either not recent or not related to cyber security. For example, data sets from the widely used Loghub repository \cite{loghub} are not cyber security-centric, with several of them more than 15 years old.

The current paper addresses the aforementioned research gaps and proposes an unsupervised LLM-based template mining approach called LLM-TD for security event logs. In the paper, we have focused on the analysis of security related Linux \emph{syslog} event logs, and the main contributions of the paper can be summarized as follows: 

\begin{itemize}

\item Whereas several existing works have utilized public LLMs \cite{lilac, divlog, ulog}, security event logs usually contain sensitive data that can not be shared with external service providers. For this reason, the current paper studies the performance of LLM-TD with several small local LLMs.

\item The paper uses five Linux security \emph{syslog} data sets for evaluating the performance of LLM-TD, and releases these data sets publicly alongside with the implementation of LLM-TD (\emph{https://github.com/ristov/llm-td}).

\item The paper analyzes performance metrics that have been previously suggested for template detection algorithms \cite{zhu, khan, logram}, and proposes several heuristic principles for addressing the shortcomings of existing metrics.

\item Finally, the paper highlights the advantages of LLM-based template detection over traditional algorithms, and discusses the importance of qualitative assessment of all identified templates.
{
In particular, LLMs have the ability to infer correct templates from insufficient event log data and identify character patterns inside log message words. Furthermore, they are able to discover previously unknown knowledge from event logs, making the qualitative analysis of both correct and incorrect templates a relevant task.
}
    
\end{itemize}

The remainder of this paper is organized as follows -- Section \ref{section_relatedwork} discusses related work, Section \ref{section_llmtd} describes the LLM-TD method, Section \ref{section_evaluation} presents its evaluation, Section \ref{section_discussion} discusses the evaluation results, and Section \ref{section_future_work} concludes the paper.

\section{Related Work}\label{section_relatedwork}

Over the last two decades, several traditional template detection algorithms have been proposed. Traditional algorithms are generally unsupervised and assume that each event log message consists of tokens or \emph{words} separated by delimiters (usually whitespace characters). The algorithms identify the words of variable nature (henceforth called \emph{variable words}), which allows the derivation of templates consisting of wildcards \wildcard{} and \emph{constant words}.

SLCT \cite{slct} begins the template mining process by detecting frequent words from event log messages, continuing with creating cluster candidates based on the combinations of frequent words. Each candidate is represented by a template, and candidates that occur for at least \emph{s} messages (\emph{s} is a user-given threshold) are selected as clusters, with their templates being reported to the end user. 

LogCluster \cite{logcluster} is a further development of SLCT that employs an advanced cluster candidate generation procedure. That procedure allows us to address the shortcomings of many other algorithms discussed in Section \ref{section_introduction} -- templates detected by LogCluster can represent similar messages that contain different numbers of words. LFA \cite{lfa} begins with identifying frequent words like SLCT and LogCluster and builds templates around frequent words that have similar occurrence times in event log messages.

LenMa \cite{lenma} relies on the following observation -- the constant words of the template have the same length (number of characters) in all messages matching that template, whereas words matching the wildcards do not have that property. Following that observation, LenMa converts event log messages into word length vectors, assigning messages with the same number of words into the same cluster based on the cosine similarity of their word length vectors. Each cluster has a template representative which is reported to the end user. 

LKE \cite{lke} employs 
expert rules for removing variable parts from event log messages, and messages are then clustered based on their edit distance, deriving a template for each cluster. AEL \cite{ael} uses heuristic rules for identifying variable words in event log messages, dividing messages into bins based on the number of constant and variable words, and creating one or more templates for each bin. LogMine \cite{logmine} first clusters event log messages by measuring their similarity in terms of common words and then generates a template for each cluster. These steps can be applied iteratively to produce more general templates from more specific ones.

IPLoM \cite{iplom} is a divisive clustering algorithm which first splits the event log into clusters by the number of words in log messages. For each cluster, a word position is then identified with the smallest number of unique words, splitting the cluster by these words. Resulting clusters are split further by associations between word pairs. After that step, a template is derived for each cluster. 

LogSig \cite{logsig} converts each log message into word pairs and finds an optimal partition of messages into clusters, trying to maximize the number of common word pairs for messages in the same cluster. Spell \cite{spell} divides the event log messages based on the longest common subsequences (LCSs) of words, creating a new LCS object for a message if it is dissimilar to existing objects or merging it with the most similar object. Each LCS object contains information about the locations of variable words in log messages corresponding to that object. After all messages have been processed, templates that are based on LCS objects are reported to the end user.

Logram \cite{logram} uses \emph{n}-gram analysis for template detection, where each \emph{n}-gram is a sequence of \emph{n} words extracted from a log message. Logram begins with creating a dictionary of 2-grams and 3-grams, which holds their occurrence times. For template detection, Logram relies on the following observation -- \emph{n}-grams of constant words are frequent, while \emph{n}-grams containing variable words are likely infrequent. To detect variable words, Logram first identifies infrequent 3-grams in event log messages and then analyzes infrequent 2-grams generated from infrequent 3-grams of the log message.

Drain \cite{drain} employs a {tree structure called} parse tree with a user-specified depth where templates are created and updated in leaf nodes. 
{
For all log messages with the same number of words, a separate child node is created under the root node of the parse tree. The nodes at the following levels contain message words and wildcards, allowing to search for a leaf node with a list of relevant templates for each log message.
}
The similarity score between the message and the template is calculated to select the most suitable template for a log message, which has to exceed a user-given similarity threshold{, and the selected template is updated (if no suitable template is found, a new template created).}

SHISO \cite{shiso} uses a tree structure for clustering log messages on the fly, and \emph{n}-gram analysis for adjusting templates during log message processing. MoLFI \cite{molfi} employs a genetic algorithm for generating templates from event log messages.

Unlike previously described traditional algorithms, LLM-based template mining approaches are generally supervised. LogPPT \cite{logppt} harnesses the Adaptive Random Sampling algorithm for selecting training data points from preprocessed event log data. Training data is then labeled with ground truth templates by the human analyst and used to \emph{fine-tune} (train) a pretrained RoBERTa language model that will be applied for template detection. The authors evaluated the model performance on 2,000-message labeled event logs from the Loghub repository.

LLMParser \cite{llmparser} employs four pretrained open-source LLMs fine-tuned with labeled examples selected from preprocessed event log data. For example selection, Mean Shift algorithm is used to cluster the preprocessed event log, and examples are sampled from detected clusters. The performance of models was measured on event logs of 2,000 messages from the Loghub repository. The authors also evaluated the performance of \emph{in-context learning}, where instead of fine-tuning, a \emph{prompt} (instructions in natural language) is presented to an LLM on how to address the template mining task {(please see Figure \ref{llm-figure} in Section \ref{section_llmtd} for an example prompt)}. According to the results in \cite{llmparser}, fine-tuning yielded higher performance than in-context learning. {As discussed in \cite{llmparser}, used LLMs accepted prompts of limited size, and insufficient log parsing instructions with a very few examples had a negative impact on performance. In contrast, fine-tuning enabled the model to learn from a larger number of more diverse examples, yielding a better performance.}

DivLog \cite{divlog} uses in-context learning for template detection, harnessing the public GPT-3 LLM through the OpenAI interface. The algorithm samples a subset of event log data with the Determinantal Point Process algorithm, and selected log messages are labeled by the human analyst. To detect a template for the log message, DivLog builds a prompt of similar message-template pairs selected from the labeled data so that LLM can learn from the provided examples to find the right template for the given message. The performance of DivLog was evaluated on 2,000-message event logs from Loghub.

LILAC \cite{lilac} uses in-context learning with the public ChatGPT LLM through the OpenAI interface, where the human analyst needs to label a subset of event log messages for prompt-building purposes. For selecting messages for labeling, a hierarchical clustering algorithm is applied for the event log, sampling messages from different clusters based on a quota distribution method. When LILAC needs to detect a template for the log message, it first checks its template cache for a suitable template in order to minimize queries to LLM. Such a design allows to apply LILAC for processing large event logs, and the authors evaluated the performance of LILAC on full-size event logs (i.e., not 2,000-message event log chunks as in several other works \cite{logppt, llmparser, divlog}) from Loghub.

Similarly to LILAC, {Lunar} \cite{ulog} employs ChatGPT LLM through OpenAI interface.
Unlike previously discussed LLM-based algorithms which are supervised \cite{logppt, llmparser, divlog, lilac}, {Lunar} is an unsupervised method and does not rely on example log messages labeled by a human analyst with ground truth templates. Whereas other approaches select specific labeled log message examples for building prompts, so that a template could be detected for \emph{one log message} according to labeled examples, {Lunar} employs a different heuristic -- if \emph{several similar log messages} are submitted to LLM, LLM is able to find a common template for them without labeled examples. Ideally, log message groups submitted to LLM should match the same template, while being diverse enough so that LLM can distinguish variable parts from constant parts in log messages. For example, in the first message group in Figure \ref{example-messages}, all variable parts of two log messages are different, which eases the template detection task for LLM.

{Lunar} employs a hierarchical log clustering algorithm to detect groups of similar log messages. First, like many traditional algorithms, {Lunar} assumes that log messages matching the same template contain the same number of words, and the event log is thus split into buckets by the number of words in log messages. By the design of {Lunar}, the further processing of each bucket is independent of other buckets, which allows the execution of these tasks in parallel. In each bucket, top-\emph{k} frequent words are identified, creating clusters for log messages that share the same frequent words, and merging clusters with too few log messages based on top-\emph{(k-1)} frequent words. To select message groups for LLM, {Lunar} samples so-called log contrastive units (LCUs) from each cluster, so that log messages in LCUs would be diverse enough. If a template is successfully detected for an LCU, all messages matching this template are removed from buckets. The sampling of LCUs from buckets and querying a common template for them from LLM continues until all buckets are empty.

\section{Unsupervised LLM-based Template Detection}\label{section_llmtd}

In this section, we will describe an unsupervised LLM-based template detection method (henceforth called LLM-TD) that has been motivated by research gaps in existing literature. As discussed in Section \ref{section_relatedwork}, unsupervised LLM-based template detection has received little attention, apart from a recent work proposing {Lunar} \cite{ulog}. However, {Lunar} suffers from a design limitation discussed in Section \ref{section_introduction}, assuming that log messages matching the same template have the same number of words (see Section \ref{section_relatedwork} for more details).
{
Because unsupervised template detection does not need a larger pool of log message examples labeled with ground truth templates, it allows to eliminate the labeling workload of human analysts, and LLM-TD is therefore employing this template detection paradigm.
}

Secondly, several existing works \cite{ulog, lilac, divlog} employ public LLMs through the OpenAI interface, which implies submitting event log data to an external service provider. However, security logs are likely to contain highly sensitive information (e.g., information about user accounts and authentication methods in an organization), and sharing it with external parties is regarded as a violation of security policies.

Whereas local LLMs allow processing event logs without external service providers, such LLMs are known to be less performant than public LLMs. For this reason, their use has been primarily considered with fine-tuning \cite{logppt, llmparser}, which, according to a recent study, yields better performance than in-context learning \cite{llmparser}. However, we argue that fine-tuning requires a considerable amount of machine learning expertise, which is not always present in an organization.

\begin{figure*}[thbp]
\centerline{\includegraphics[width=1\textwidth]{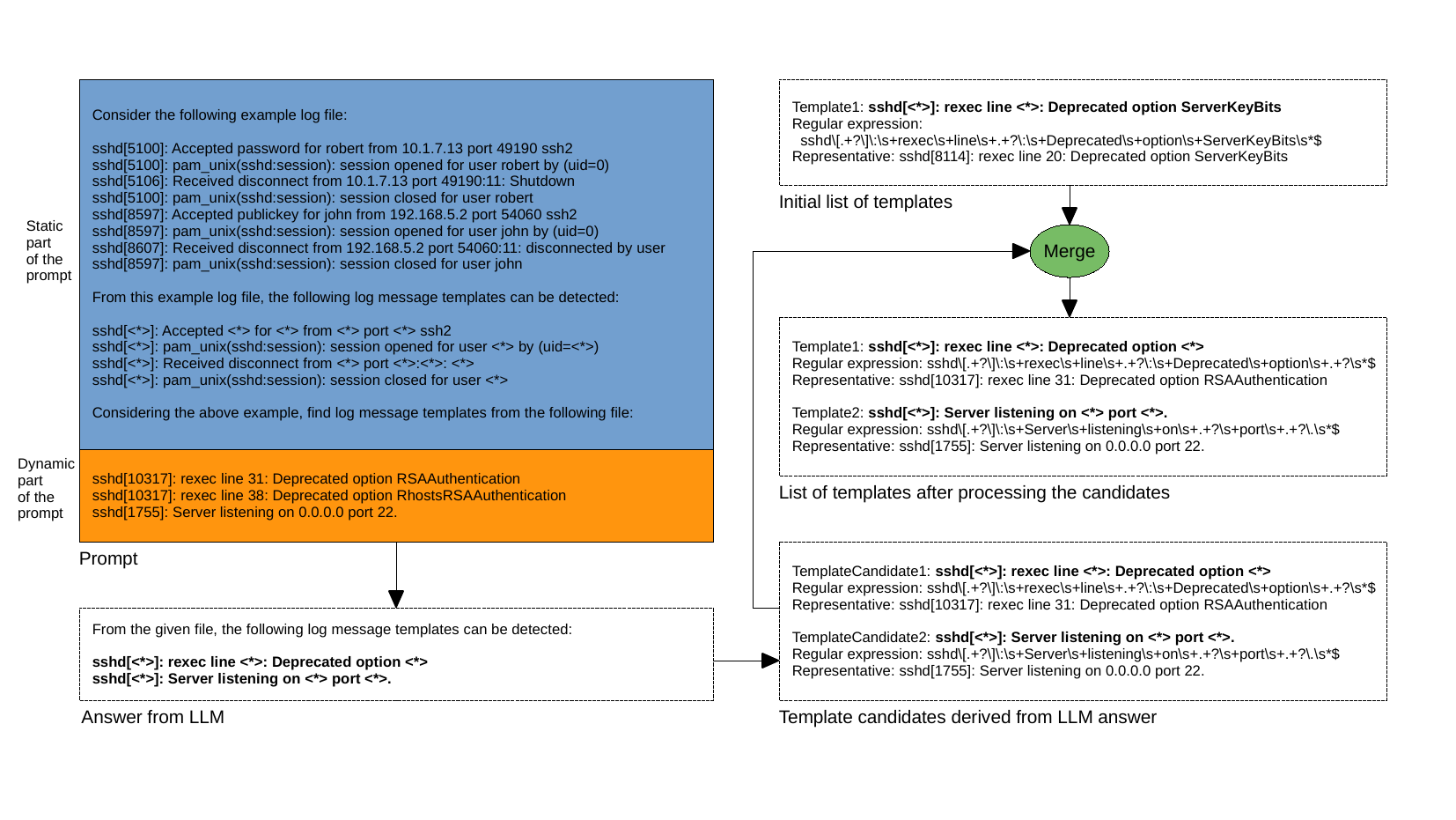}}
\caption{Interaction between LLM-TD and the underlying LLM (\emph{QueryLLMforTemplates} and \emph{Merge} procedures from Algorithm \ref{llm-td})}
\label{llm-figure}
\end{figure*}

For the reasons above, this paper explores the use of local LLMs for security event log analysis, and the LLM-TD method employs an unsupervised in-context learning paradigm for template detection. Whereas other methods attempt to detect \emph{one} template with each LLM query (i.e., templates are detected one-by-one), LLM-TD uses a different approach, which is able to identify \emph{several} templates with one query. In order to achieve that, LLM-TD utilizes a prompt that formulates the task of a template detection for a small event log which contains messages matching \emph{several templates}, not for one message or a group of similar messages which match \emph{one template only} {(Figure \ref{llm-figure} depicts the prompt used by LLM-TD)}.

When compared to {Lunar}, which is another unsupervised LLM-based algorithm (see Section \ref{section_relatedwork}), both LLM-TD and {Lunar} process a batch of several log messages with one LLM query because it can ease the unsupervised template detection task for LLMs \cite{ulog}. However, {Lunar} harnesses a dedicated multi-stage selection method for ensuring the similarity of messages within a batch, so that a single template can be detected for the batch. In contrast, as LLM-TD does not aim to detect one template {only} which must cover the entire batch, it can process less similar messages with one LLM query, and thus does not need a complex method for creating message batches. {Note that LLM-TD does not specify any constraints to the number of detected templates in LLM query, and LLM is free to report back \emph{any} number of templates (e.g., one, two, five, etc.) as it sees most appropriate for the submitted batch.}

As pointed out in \cite{ulog}, unsupervised detection of templates is complicated if too diverse log messages are submitted to LLM. Following that observation, LLM-TD divides the event log data into different parts by applications (e.g., sshd, Suricata) that have produced the log messages, where each part is processed independently. Similarly to {Lunar}, dividing the event log data allows for parallel processing but does not restrict templates to match the messages with the same number of words (unlike {Lunar}). Having each part containing messages from the same application leads to much less diversity among log messages and fewer templates that need detection. 

Also, in the case of Linux \emph{syslog}-based security logs (the primary focus of this paper), the application name can be retrieved from the \emph{syslog} tag field which is defined by the BSD \emph{syslog} protocol \cite{bsd-syslog}. Since the purpose of LLM-TD is the analysis of unstructured log message data, LLM-TD implementation will not consider timestamp and hostname fields of \emph{syslog} messages, targeting the \emph{syslog} tag field which is followed by a free-form textual message string (Figure \ref{example-messages} depicts such \emph{syslog} message examples).
{Note that the location of timestamp, hostname, \emph{syslog} tag (application name), and message string fields are well documented in the \emph{syslog} protocol standard \cite{bsd-syslog}, and therefore the identification of these fields does not assume detailed domain knowledge about the individual message formats of logging applications.}

\begin{algorithm}
\caption{\emph{Event Log Analysis with LLM-TD}}\label{llm-td}
\begin{algorithmic}[1]
\Require $L = (l_1, ..., l_n)$ -- event log of \emph{n} messages,
\Statex ~~~~~$k$ -- batch size
\Ensure $T$ -- templates, $U$ -- uncovered messages,
\Statex ~~~~~~~$V$ -- duplicate templates

\State $T \gets \varnothing $ 
\State $B \gets ()$

\State \textbf{for each} $l \in L$
\State ~~~\textbf{if} $\exists t \in T$, so that $l$ matches $t$ \textbf{then} next loop iteration
\State ~~~\textbf{append} $l$ \textbf{to} $B$
\State ~~~\textbf{if} $|B| < k$ \textbf{then} next loop iteration
\State ~~~$X \gets$ QueryLLMforTemplates($B$)
\State ~~~Merge(T, X)
\State ~~~$B \gets ()$

\State \textbf{if} $|B| > 0$ \textbf{then}
\State ~~~$X \gets$ QueryLLMforTemplates($B$)
\State ~~~Merge(T, X)

\State $U \gets \varnothing $
\State $Y \gets \varnothing $

\State \textbf{for each} $l \in L$
\State ~~~$M \gets \{t~|~t \in T,~l~matches~t\}$
\State ~~~\textbf{if} $|M| = 0$ \textbf{then} $U \gets U \cup \{~l~\}$
\State ~~~\textbf{if} $|M| = 1$ \textbf{then} $Y \gets Y \cup M$

\State $V \gets T \setminus Y$

\State \textbf{return} T, U, V

\end{algorithmic}
\end{algorithm}

After event log data has been divided into parts by the name of the logging application, Algorithm \ref{llm-td} is applied to each part (that is, an event log of an individual application). LLM-TD algorithm makes two passes over the event log, where the first pass (lines 1--12 in Algorithm \ref{llm-td}) involves template detection, and the second pass (lines 13--19 in Algorithm \ref{llm-td}) the identification of duplicate templates and log messages not covered by templates, allowing the human expert to analyze challenging messages and potentially redundant templates manually.

During the first pass, log messages are processed sequentially, submitting them to LLM in batches of \emph{k} messages. If a log message matches an already detected template in the template set \emph{T}, it is skipped (line 4 in Algorithm \ref{llm-td}) for decreasing the number of computationally expensive queries to LLM (for matching the message with the template, the regular expression representation of the template is used). After the batch of \emph{k} messages has been assembled for processing by LLM, the batch is included in a prompt that is illustrated in Figure \ref{llm-figure}. After the prompt has been created, it is used for querying the LLM (line 7 in Algorithm \ref{llm-td}).

The prompt consists of a static and dynamic part, with the first containing a hardcoded example of how to detect multiple templates from a small log file, whereas the dynamic part contains the message batch that needs processing. As a response to this prompt, LLM provides an answer in natural language (see Figure \ref{llm-figure}). In order to pick up template definitions from LLM output, it is processed with regular expression-based filters, and for each definition, a corresponding regular expression is created automatically (see the template candidate box in Figure \ref{llm-figure}). To check if the template definition is valid for the given message batch, its regular expression is matched against the messages, and the template is discarded if there is no match. Otherwise, the template is regarded as a \emph{template candidate} and the first matching message in the batch is considered its \emph{representative}.

After the candidates have been identified, they are merged with the set of templates \emph{T} (line 8 in Algorithm \ref{llm-td}). The process of merging has been illustrated in Figure \ref{llm-figure}. Firstly, candidate definitions already present in \emph{T} are dropped, adding the remaining definitions to \emph{T}. After that, all definitions in \emph{T} are compared by matching their representatives in order to detect and drop definitions that are more specific than others. That step is taken to increase the computational efficiency since templates in \emph{T} are employed as filters for event log data (line 4 in Algorithm \ref{llm-td}). For example, in Figure \ref{llm-figure}, the representative of the template \emph{sshd[\wildcard{}]: rexec line \wildcard{}: Deprecated option ServerKeyBits} matches the candidate template \emph{sshd[\wildcard{}]: rexec line \wildcard{}: Deprecated option \wildcard{}}. On the other hand, the representative of the candidate template does not match the template \emph{sshd[\wildcard{}]: rexec line \wildcard{}: Deprecated option ServerKeyBits}. Therefore, the template \emph{sshd[\wildcard{}]: rexec line \wildcard{}: Deprecated option ServerKeyBits} will be dropped from \emph{T}.

After the first pass over the event log is complete, LLM-TD checks for the presence of an incomplete batch with less than \emph{k} messages and, if found, submits it to LLM (lines 10--12 in Algorithm \ref{llm-td}). During the second data pass (lines 13--19 in Algorithm \ref{llm-td}), event log messages not matching any templates from \emph{T} are identified, assigning them to set \emph{U}. In addition, duplicate templates are highlighted by assigning them to set \emph{V}. A template is regarded as a duplicate if no event log message is matched by this template alone. For instance, if all successful SSH public key-based logins originate from the host 10.1.1.1, all such messages always match both of the following templates, which makes these templates duplicates:

\emph{sshd[\wildcard{}]: Accepted publickey for \wildcard{} from \wildcard{} port \wildcard{} ssh2} 

\emph{sshd[\wildcard{}]: Accepted \wildcard{} for \wildcard{} from 10.1.1.1 port \wildcard{} ssh2} 

Therefore, highlighting them will help the human expert to derive a correct template from them manually:

\emph{sshd[\wildcard{}]: Accepted \wildcard{} for \wildcard{} from \wildcard{} port \wildcard{} ssh2}

\section{Evaluation}\label{section_evaluation}

\subsection{Experiment Environment and Data Sets}\label{section_experiment_env}

For conducting the experiments with LLM-TD on Linux security \emph{syslog} files, we created an implementation of LLM-TD in Perl, with the \emph{QueryLLMforTemplates} procedure from Algorithm \ref{llm-td} being implemented in Python using the LangChain framework. For local LLMs, we selected {three} local LLMs -- OpenChat \cite{openchat}, Mistral \cite{mistral}, and Wizardlm2 \cite{wizardlm2}. In the case of some benchmarks \cite{openchat}, OpenChat has been reported to offer comparable performance to ChatGPT, which has been employed by several previous works for template detection \cite{lilac, ulog}. Also, Mistral and Wizardlm2 are state-of-the-art LLMs from Mistral AI and Microsoft AI, respectively, which we found to work well for template detection tasks. 
{
We also considered a number of other local LLMs for general purposes (Llama2, Llama3, Mixtral, Phi, and Phi3) and more specific goals such as code analysis and mathematical reasoning (Codellama, Codeup, and Wizard-math), but they featured lower runtime and template detection performance than the selected three LLMs.
}

For all three models, we employed their 7B implementations through the Ollama framework, which was installed on a Rocky Linux 9 workstation with NVIDIA GeForce RTX 2080 Ti video card, Intel Core i9-10900X CPU, and 128GB of memory. Therefore, the evaluation allowed us to establish how well small LLMs with parameter size 7B would be suited for mining templates from security event logs using the computing power of only one workstation.

In order to find how well LLM-TD with small LLMs is performing in comparison to traditional algorithms, we selected Drain \cite{drain} as a baseline because it was identified as the most performant traditional algorithm in the study by Zhu et al. \cite{zhu}. Also, Drain has been used as a baseline in several recent studies on LLM-based template detection \cite{logppt, divlog, lilac, ulog, llmparser}. We were not able to include {Lunar} (the only existing unsupervised LLM-based algorithm) in the comparison since its implementation is not publicly available.

\begin{table}[htbp]
\caption{Event log data sets}
\begin{center}
\begin{tabular}{c c c c}
\hline
\textbf{Data set} & \textbf{Number of} & \textbf{Number of} & \textbf{Number of} \\
& \textbf{log messages} & \textbf{ground truth} & \textbf{logging hosts} \\
& & \textbf{templates} & \\
\hline
sshd & 632,677 & 36 & 37 \\
su & 2354 & 13 & 28 \\
suricata & 999,622 & 6 & 1 \\
snmpd & 108,935 & 7 & 6 \\
apache & 63,947 & 7 & 1 \\
\hline
\end{tabular}
\label{datasets}
\end{center}
\end{table}

Five medium-sized Linux \emph{syslog} data sets labeled with ground truth templates were used for evaluation, with the data sets being described in Table \ref{datasets}. The medium size of data sets eased the task of qualitative analysis of detected templates (the importance of this task is clarified in {Section \ref{section_results_erroranalysis}}). 

The data sets covered different areas of network and system security: network intrusion detection {(\emph{suricata} data set)}, authentication and access over SSH and SNMP protocols {(\emph{sshd} and \emph{snmpd} data sets, respectively)}, local authentication {(\emph{su} data set)}, and web application code quality issues {(\emph{apache} data set)}. Each data set contained messages from one application only -- \emph{sshd} daemon messages from 37 hosts, \emph{su} authentication messages from 28 hosts, Suricata network IDS messages from one host, \emph{snmpd} daemon messages that reflected SNMP-based access to 6 hosts, and Apache web server messages about code quality issues of PHP applications running on one host.

{
All data sets were collected in a large academic institution from a production environment (i.e., data sets were not created in a lab environment through some experiments, but reflect real-life log data of an organizational IT system).
The \emph{sshd}, \emph{su}, \emph{snmpd}, and \emph{apache} data sets were collected during 35 days, and the \emph{suricata} data set during 5 days. The \emph{suricata} data set originated from a Suricata IDS which was monitoring an external network perimeter of the entire organization and routinely processing 2--3 Gbit/s of network traffic. The \emph{apache} data set was collected from publicly accessible organizational web portal, whereas remaining data sets originated from system daemons of critical Linux hosts.
}

During the experiments, we ran Drain with similarity threshold and {parse} tree depth parameters set to 0.39 and 6, respectively (the same settings were used for \emph{syslog} data in the comparative study by Zhu et al. \cite{zhu}). 
{
Due to the unsupervised nature of the LLM-TD algorithm, it was used with the \emph{same} static part of the LLM prompt (the blue part of the prompt in Figure \ref{llm-figure} from Section \ref{section_llmtd}) for \emph{all} data sets from Table \ref{datasets}.\footnote{
In contrast, with supervised methods the template detection examples (see the blue part of the prompt in Figure \ref{llm-figure}) would no longer be static, but dynamically created for each log message batch, selecting the most appropriate examples from a larger pool previously developed by human experts. With unsupervised LLM-TD, the creation of such a large pool of labeled examples is not needed.}
}

\subsection{Performance Metrics}\label{section_metrics}

\begin{figure}[htbp]
\centerline{\includegraphics[width=0.5\textwidth]{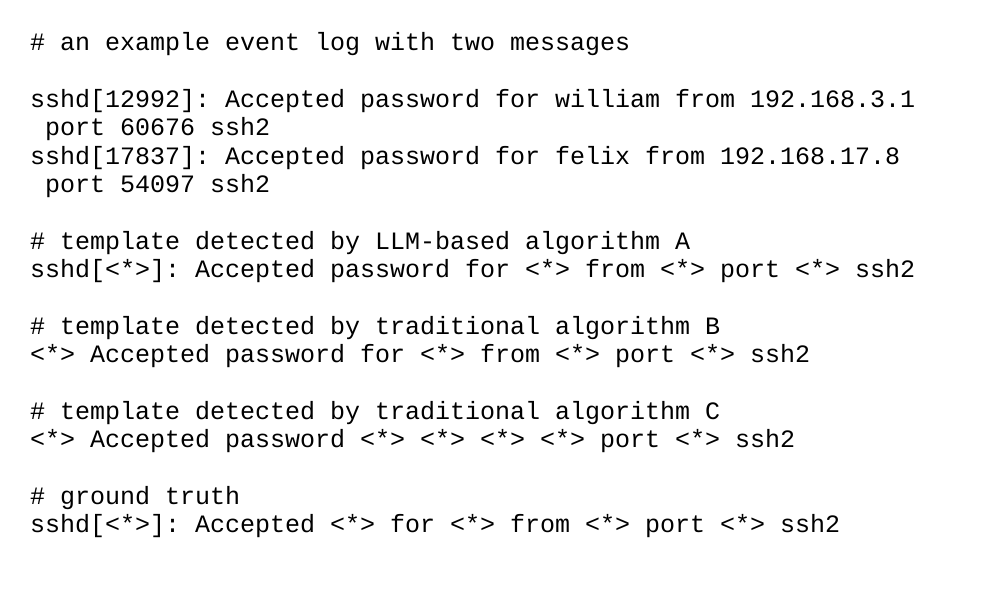}}
\caption{An example event log with detected templates and ground truth}
\label{metrics-examples}
\end{figure}

In the research literature, several different metrics have been proposed for assessing the performance of template detection algorithms, and a recent study by Khan et al. \cite{khan} provides a detailed treatment of these metrics. To support the discussion in this section, an example event log in Figure \ref{metrics-examples} is provided, which assumes that the same template describes all messages in the log.

The \emph{grouping accuracy} \cite{zhu} is a metric that measures the proportion of correctly parsed messages in the event log. A message is regarded as correctly parsed if its template corresponds to the same group of messages as defined by the ground truth. For example, in Figure \ref{metrics-examples} algorithms \emph{A}, \emph{B}, and \emph{C} have created one template for two messages found in the event log, and have thus reached the same grouping as ground truth. However, as pointed out in recent studies \cite{gasimov, khan, logram}, grouping accuracy does not reflect if detected templates are accurately reflecting constant and variable parts of log messages (e.g., the result from algorithm \emph{C} in Figure \ref{metrics-examples} reports several constant parts as variables).

The \emph{parsing accuracy} \cite{logram} measures the proportion of correctly parsed messages in the event log and regards the message as correctly parsed if its template identifies all constant and variable parts correctly. In \cite{logram}, Dai et al. conducted the manual analysis of messages and templates for establishing the parsing accuracy. Considering the ground truth, no algorithm in Figure \ref{metrics-examples} has parsed the messages correctly.

Grouping and parsing accuracy have one weakness -- if only a small part of templates cover the majority of the event log, detecting these templates correctly will yield high accuracy, although they represent a small minority among templates. To address this issue, Khan et al. \cite{khan} have proposed to consider detected templates with equal weight and have suggested the concept of \emph{template accuracy}. According to this concept, a detected template \emph{t} is correctly identified if all log messages that the template \emph{t} matches correspond to 
a single (i.e., not more than one)
ground truth template \emph{v}, so that templates \emph{t} and \emph{v} are identical. As seen from Figure \ref{metrics-examples}, none of the three detected templates is correctly identified since the messages they match correspond to a different ground truth template.

However, existing metrics face several challenges illustrated by Figure \ref{metrics-examples}. Firstly, although in Figure \ref{metrics-examples} the word \emph{password} is actually a variable part in log messages and denotes the authentication method (see also Figure \ref{example-messages} in Section \ref{section_introduction}), there are no other authentication method keywords present in log data. This complicates the identification of the ground truth template, and traditional algorithms are generally unable to handle this task. Therefore, a performance metric should take into account the nature of event log data, and if some \wildcard{} in the ground truth template is always matching the same string \emph{s} in event log data, a template where \emph{s} appears in place of that \wildcard{} should be regarded as correctly identified.
Following that logic, the template detected by algorithm \emph{A} is correct since the second \wildcard{} in the ground truth template always matches the same string (\emph{password}) as reported in the detected template. In the remainder of the paper, we use this principle (henceforth called P1) to decide if a detected template is correct.

A simplistic workaround to the aforementioned problem would be to adjust the ground truth template according to the event log data (e.g., in Figure \ref{metrics-examples} that would involve using the template from algorithm \emph{A} as ground truth). Whereas that approach would indeed be appropriate for traditional algorithms, LLM-based solutions are, in some cases, able to correctly infer \wildcard{} even from constant data. In the rest of the paper, we call this task \emph{inferring the ground truth template from insufficient event log data}. Therefore, adjusting the ground truth templates would not allow to adequately assess the capabilities of LLM-based algorithms for addressing this task. To summarize, the P1 principle allows the evaluation of traditional and LLM-based algorithms, taking into account the limitations of traditional algorithms. However, such evaluation should also qualitatively assess which correct templates are more similar to the ground truth templates for the given event log data.

The second challenge to the existing metrics comes from another limitation of traditional algorithms. Namely, most algorithms assume that each event log message consists of words that are separated by delimiters (usually whitespace characters), and no attempt is made to analyze the content of words by individual characters. As a result, the words \emph{sshd[12992]:} and \emph{sshd[17837]:} from Figure \ref{metrics-examples} are identified as variables by many algorithms and reported as \wildcard{}, with a template from algorithm \emph{B} illustrating that problem.

For a workaround, two approaches have been suggested, with the first being the preprocessing of event log data (e.g., by replacing variable substrings in words with \wildcard{}) \cite{zhu}. However, as discussed in Section \ref{section_introduction}, preprocessing requires detailed knowledge of message formats. Also, Khan et al. have proposed the modification of ground truth templates with heuristic rules \cite{khan}, which can replace a template word with \wildcard{} if that word contains \wildcard{} as a substring. For example, \emph{sshd[\wildcard{}]:} would be converted to \wildcard{} in the ground truth template from Figure \ref{metrics-examples}, making it identical to the template detected by algorithm \emph{B}. Similarly to the workaround for the first challenge to performance metrics, it would be appropriate for traditional algorithms. However, LLM-based algorithms are able to identify variable parts inside words, and the adjustment of ground truth would downplay the strengths of LLM-based methods. 

As a solution for this problem, we have used the following principle (henceforth called P2), which takes into account the limitations of traditional algorithms. If some word in the ground truth template contains \wildcard{} as a substring, a detected template where \wildcard{} appears in place of that word is regarded correct, provided that this template does not match more messages than the ground truth template 
(i.e., replacing the entire word with \wildcard{} does not have any impact on the given event log data).
Note that we have applied principle P2 \emph{only to traditional algorithms} (i.e., Drain) in this paper. According to this principle, the template detected by algorithm \emph{B} in Figure \ref{metrics-examples} is correct.

In addition to using principles P1 and P2 for assessing the correctness of an individual template, this paper follows the example of previous studies \cite{lilac, ulog} and employs the F1-score metric (harmonic mean of precision and recall) for measuring the performance of algorithms. \emph{Precision} is defined as the ratio of correctly identified templates and all detected templates, whereas \emph{recall} is the ratio of correctly identified templates and all ground truth templates. Also, F1-score is defined as:

\begin{equation}
F1 = 2 * \frac{precision * recall}{precision + recall}
\end{equation}

\subsection{Results}\label{section_results}

{
This section presents the evaluation results for four selected approaches (LLM-TD with three LLMs and Drain). Section \ref{section_results_exectime} discusses the execution time of evaluated approaches and the impact of log message batch size to the runtime of LLM-TD. Section \ref{section_results_tempdetection} covers the template detection performance and Section \ref{section_results_erroranalysis} provides the analysis of incorrectly detected templates.
}

{\subsubsection{Execution Time}\label{section_results_exectime}}

For measuring the execution time and performance of four different template detection algorithms described in Section \ref{section_experiment_env} (Drain and LLM-TD with OpenChat, Mistral, and Wizardlm2), we applied them to data sets from Table \ref{datasets} 10 times. Repeated execution allowed us to estimate the execution time of algorithms better, and since an LLM can produce different output for the same input data, we were able to estimate how the number of detected templates varies for LLM-based approaches. Since the implementations of Drain and LLM-TD are single-threaded, the number of CPUs of the experiment machine did not influence the execution time. Table \ref{results-runtime} shows the results for 10 experiment iterations. Note that in Table \ref{results-runtime} and the remainder of the paper, we have used the name of an LLM to denote LLM-TD using that LLM (e.g., OpenChat denotes LLM-TD with OpenChat). 
{
Also, if an algorithm was not able to process a data set within 10 hours, respective experiment was terminated and corresponding cell in Table \ref{results-runtime} contains the '-' character.\footnote{This allowed to limit the overall duration of experiments to 12 weeks, because 10 iterations would take at most 100 hours, and evaluating four algorithms on five data sets would thus require at most 2000 hours, which is about 2000 / 24 $\approx$ 83.33 days $\approx$ 12 weeks.}
}

\begin{table*}[htbp]
\caption{Execution time of template detection methods}
\begin{center}
\begin{tabular}{c c c c c c|c c c c c}
\hline
\textbf{Method} & \multicolumn{5}{c}{\textbf{Average runtime in seconds}} & \multicolumn{5}{c}{\textbf{Number of detected templates}} \\
& \multicolumn{5}{c}{\textbf{(standard deviation in parentheses)}} & \multicolumn{5}{c}{\textbf{(the number of uncovered messages in parentheses)}} \\
\cline{2-11} 
& \textbf{\textit{sshd}} & \textbf{\textit{su}} & \textbf{\textit{suricata}} & \textbf{\textit{snmpd}} & \textbf{\textit{apache}} & \textbf{\textit{sshd}} & \textbf{\textit{su}} & \textbf{\textit{suricata}} & \textbf{\textit{snmpd}} & \textbf{\textit{apache}} \\
\hline
OpenChat & 175.136 & 15.494 & 1195.762 & 7.086 & 6.984 & 29 & 1 & 22 & 7 & 7 \\
\emph{k}=2 & (0.334) & (0.017) & (2.64) & (0.031) & (0.021) & (189) & (0) & (249) & (0) & (0) \\
\hline
OpenChat & 77.413 & 8.511 &  443.732 & 6.274 & 6.742 & 28 & 1 & 43 & 7 & 7 \\
\emph{k}=5 & (0.265) & (0.011) & (1.214) & (0.032) & (0.018) & (148) & (0) & (58) & (0) & (0) \\
\hline
OpenChat & 48.644 & 12.645 &  90.544 & 5.868 & 12.609 & 31 & 13 & 15 & 7 & 7 \\
\emph{k}=10 & (0.44) & (0.019) & (0.737) & (0.038) & (0.091) & (4) & (1) & (2) & (0) & (0) \\
\hline
OpenChat & 85.694 & 13.953 & - & 6.395 & 6.364 & 27 & 13 & - & 7 & 7 \\
\emph{k}=20 & (0.395) & (0.011) & & (0.785) & (0.017) & (223) & (0) & & (0) & (0) \\
\hline
Mistral & 309.399 & 21.925 & 2542.645 & 7154.681 & 3745.85 & 32 & 2 & 2 & 5 & 2 \\
\emph{k}=2 & (0.505) & (0.008) & (0.71) & (4.854) & (0.835) & (43) & (0) & (102) & (0) & (0) \\
\hline
Mistral & 166.244 & 44.358 & - & 25.052 & 5.262 & 23 & 10 & - & 7 & 1 \\
\emph{k}=5 & (1.083) & (0.022) &  & (1.073) & (0.004) & (54) & (39) &  & (4) & (0) \\
\hline
Mistral & 59.088 & 17.51 & - & 26.402 & 8.088 & 25 & 10 & - & 1 & 2 \\
\emph{k}=10 & (0.433) & (0.004) &  & (0.071) & (0.01) & (65) & (6) &  & (0) & (0) \\
\hline
Mistral & 7544.625 & 13.787 & - & 10.737 & 7.32 & 22 & 10 & - & 5 & 2 \\
\emph{k}=20 & (2.382) & (0.015) &  & (0.785) & (0.008) & (204) & (6) &  & (12) & (0) \\
\hline
Wizardlm2 & - & 3176.84 & 8540.346 & 873.894 & 7771.931 & - & 75 & 1 & 8 & 4 \\
\emph{k}=2 & & (1.084) & (3.212) & (0.158) & (5.215) & & (1163) & (0) & (10) & (2) \\
\hline
Wizardlm2 & 5400.69 & 1351.209 & - & 1042.132 & 8119.098 & 56 & 22 & - & 6 & 8 \\
\emph{k}=5 & (1.528) & (0.155) & & (0.766) & (1.593) & (491) & (244) & & (835) & (8245) \\
\hline
Wizardlm2 & 1824.513 & 1309.906 & - & 93.4 & 324.047 & 51 & 9 & - & 6 & 7 \\
\emph{k}=10 & (0.591) & (0.162) & & (1.243) & (0.086) & (1121) & (822) & & (12) & (15) \\
\hline
Wizardlm2 & 906.329 & 654.405 & - & 186.198 & 323.856 & 20 & 20 & - & 6 & 7 \\
\emph{k}=20 & (0.737) & (0.153) & & (1.584) & (0.062) & (957) & (927) & & (12) & (15) \\
\hline
Drain & \textbf{9.904} & \textbf{0.091} & \textbf{19.696} & \textbf{1.662} & \textbf{1.065} & 44 & 16 & 80 & 8 & 4 \\
& (0.037) & (0.003) & (0.151) & (0.012) & (0.008) & (0) & (0) & (0) & (0) & (0) \\
\hline
\end{tabular}
\label{results-runtime}
\end{center}
\end{table*}

{
Our first notable finding was that although LLMs are known to be non-deterministic, the LLM-TD algorithm produced the same result for all 10 iterations of every experiment scenario. In other words, LLM-TD featured a deterministic behavior during all experiments, making it similar to traditional algorithms which are usually deterministic.

We also assessed the influence of the log message batch size (parameter \emph{k} in Algorithm \ref{llm-td}) to the speed of the LLM-TD algorithm for the following settings: \emph{k}=2, 5, 10, 20 (i.e., batch size was increased by 2--2.5 times for the next scenario). As Table \ref{results-runtime} shows, OpenChat featured short execution times of 5--15 seconds for three smaller data sets (\emph{su}, \emph{snmpd} and \emph{apache}) for all values of \emph{k}. For two larger and more challenging data sets (\emph{sshd} and \emph{suricata}), increasing the value of \emph{k} from 2 to 10 decreased the runtime, whereas increasing \emph{k} from 10 to 20 provided no further performance benefits. For example, with \emph{k}=10 the \emph{suricata} data set was processed in 1.5 minutes, while with \emph{k}=20 the runtime exceeded 10 hours.

For Mistral, only the smallest \emph{su} data set featured short execution times (13--44 seconds) for all values of \emph{k}. For the \emph{snmpd} and \emph{apache} data sets, \emph{k}=2 provided the worst runtime performance. Interestingly, for the largest \emph{suricata} data set the opposite pattern was observed, and \emph{k}=2 was the only setting which led to a runtime of less than 10 hours (about 42 minutes). However, when investigating this phenomenon, we discovered it was caused by the detection of an overly general template in early phases of event log processing, which caused the exclusion of most log data from further processing (see line 4 in Algorithm \ref{llm-td}). We will provide a detailed discussion of that issue in Section \ref{section_results_erroranalysis}. Apart from the \emph{suricata} and \emph{su} data set, increasing the value of \emph{k} until 10 provided significant performance benefits for Mistral, and with \emph{k}=10 all data sets except \emph{suricata} were processed in less than 1 minute. However, increasing \emph{k} from 10 to 20 led to a poor runtime performance of over 2 hours on the \emph{sshd} data set.

Like Mistral, Wizardlm2 was able to process the \emph{suricata} data set in 10 hours only with \emph{k}=2, and the reasons for this were the same as for Mistral (the detection of an overly general template). Apart from the \emph{sshd} data set, increasing \emph{k} from 2 to 5 provided less noticeable runtime performance benefits than observed for Mistral. However, increasing \emph{k} further to 10 decreased the runtime significantly for \emph{sshd}, \emph{snmpd}, and \emph{apache} data sets, with all data sets except \emph{suricata} being processed within 1.5--30 minutes. Increasing the value of \emph{k} from 10 to 20 improved performance on \emph{sshd} and \emph{su} data sets, while leading to a longer runtime on the \emph{snmpd} data set.

According to the analysis above, \emph{k}=10 yielded the best overall performance for LLM-TD with OpenChat and Mistral, and for Wizardlm2 both \emph{k}=10 and  \emph{k}=20 featured a good performance. Because \emph{k}=10 also yielded the best F1-scores for the majority of experiment scenarios, further experiments described in this paper were conducted with this particular setting (i.e., the log message batch size of 10).
}

As Table \ref{results-runtime} illustrates, Drain was the fastest algorithm for all data sets, having an average execution time of less than 20 seconds in all cases. {With \emph{k}=10,} OpenChat was the second fastest on most data sets, running for less than {91 seconds on all data sets}, and was the only LLM-based approach that was able to process all data sets successfully. Mistral and Wizardlm2 did not complete the analysis of the \emph{suricata} data set in 10 hours {for most settings of \emph{k}}. On the other four data sets, Wizardlm2 was {usually} the slowest approach {-- for example, with \emph{k}=10, it had an} execution time of {about} 5, {22}, and 30 minutes on \emph{apache}, \emph{su}, and \emph{sshd} data sets, respectively.

{
When investigating the causes of execution time differences of LLM-TD with different LLMs, we discovered the number of LLM queries as one of the reasons, and Table \ref{results-llm-interaction} shows the relevant data for \emph{k}=10 (for the sake of brevity, data for other values of \emph{k} has been omitted, because it followed the same pattern as data in Table \ref{results-llm-interaction}). As Table \ref{results-llm-interaction} illustrates, OpenChat and Mistral were able to process all data sets with only 3--24 queries, which involved a modest amount of total LLM query time. However, Wizardlm2 executed noticeably more queries for \emph{apache}, \emph{su}, and \emph{sshd} data sets, and that resulted in a significantly more time spent for running these queries. 

The need for a larger number of queries is connected to the fact that Wizardlm2 generated more invalid templates than other LLMs (e.g., 56--316 invalid templates for \emph{apache}, \emph{su}, and \emph{sshd} data sets, whereas OpenChat and Mistral generated 1--34 invalid templates for these data sets). As discussed in Section \ref{section_llmtd}, LLM-TD validates each template returned from LLM by checking if this template matches at least one message in the message batch which LLM had to process. Because Wizardlm2 made more mistakes in the template detection process, that process became much slower (e.g., LLM-TD was able to skip less messages in line 4 of Algorithm \ref{llm-td}), leading to longer execution times. 

Another reason for execution time differences was the response time for LLM queries. Based on the data from Table \ref{results-llm-interaction}, the average response time of OpenChat was 1.7--2.3 seconds on the \emph{sshd}, \emph{su}, \emph{snmpd}, and \emph{apache} data sets, and 2.8 seconds on the \emph{suricata} data set. For Mistral, the average response time on the \emph{sshd}, \emph{su}, \emph{snmpd}, and \emph{apache} data sets was 1.9--2.5 seconds, whereas Wizardlm2 (the slowest LLM) had the average response time of 4.1--9.3 seconds on these data sets. Therefore, a noticeably longer LLM query response time of Wizardlm2 was another factor behind its longer execution times.

Due to long runtimes of Mistral and Wizardlm2 on the \emph{suricata} data set, we were not able to execute them 10 times as for other scenarios shown in Table \ref{results-llm-interaction}, and we conducted only one experiment iteration for assessing the impact of the number of LLM queries, the average query response time, and the number of invalid templates on the execution time. Mistral produced 23,210 invalid templates during 15,700 LLM queries with an average query response time of 5.1 seconds, which led to the runtime of 22 $\frac{1}{4}$ hours (i.e., almost one day). Wizardlm2 produced 8577 invalid templates during 42,099 LLM queries with an average query response time of 9.7 seconds, having the runtime of 115 hours (i.e., almost five days). In both cases, vast majority of the runtime was spent for LLM queries. These results illustrate that a larger number of invalid templates are likely to involve a larger number of LLM queries, and when combined with longer query response times, the overall impact on the execution time is significant. It should be noted that long execution times on the \emph{suricata} data set did not lead to a good template detection performance, and the F1-scores of Mistral and Wizardlm2 remained below 0.02.
}

\begin{table*}[htbp]
\caption{Interaction with LLM (log message batch size 10)}
\begin{center}
\begin{tabular}{c c c c c c|c c c c c}
\hline
\textbf{Method} & \multicolumn{5}{c}{\textbf{Number of LLM queries}} & \multicolumn{5}{c}{\textbf{Total LLM query time}} \\
& \multicolumn{5}{c}{} & \multicolumn{5}{c}{\textbf{(standard deviation in parentheses)}} \\
\cline{2-11} 
& \textbf{\textit{sshd}} & \textbf{\textit{su}} & \textbf{\textit{suricata}} & \textbf{\textit{snmpd}} & \textbf{\textit{apache}} & \textbf{\textit{sshd}} & \textbf{\textit{su}} & \textbf{\textit{suricata}} & \textbf{\textit{snmpd}} & \textbf{\textit{apache}} \\
\hline
OpenChat & 19 & 7 & 24 & 3 & 5 & 35.562 & 12.607 & 68.383 & 4.991 & 11.633 \\
& & & & & & (0.165) & (0.019) & (0.049) & (0.029) & (0.006) \\
\hline
Mistral & 19 & 7 & - & 13 & 4 & 47.269 & 17.475 & - & 25.587 & 7.556 \\
& & & & & & (0.43) & (0.007) &  & (0.073) & (0.01) \\
\hline
Wizardlm2 & 194 & 182 & - & 16 & 78 & 1809.316 & 1309.777 & - & 92.519 & 323.523 \\
& & & & & & (0.476) & (0.161) & & (1.236) & (0.085) \\
\hline
\end{tabular}
\label{results-llm-interaction}
\end{center}
\end{table*}

{\subsubsection{Template Detection Performance}\label{section_results_tempdetection}}

{
As discussed in the previous section, log message batch size 10 (\emph{k}=10) tended to yield the best execution times and F1-scores for LLM-based approaches, and this section presents template detection results for this setting. Table \ref{results-performance} shows the template detection performance in terms of F1-scores for all evaluated methods.
}

\begin{table*}[htbp]
\caption{Performance of template detection methods (log message batch size 10)}
\begin{center}
\begin{tabular}{c c c c c c|c c c c c}
\hline
\textbf{Method} & \multicolumn{5}{c}{\textbf{Number of correctly detected templates}} & \multicolumn{5}{c}{\textbf{F1-score}} \\
& \multicolumn{5}{c}{\textbf{(number of all detected templates in parentheses)}} & \multicolumn{5}{c}{} \\
\cline{2-11} 
& \textbf{\textit{sshd}} & \textbf{\textit{su}} & \textbf{\textit{suricata}} & \textbf{\textit{snmpd}} & \textbf{\textit{apache}} & \textbf{\textit{sshd}} & \textbf{\textit{su}} & \textbf{\textit{suricata}} & \textbf{\textit{snmpd}} & \textbf{\textit{apache}} \\
\hline
OpenChat & 25 (31) & 10 (13) & 3 (15) & 7 (7) & 5 (7) & \textbf{0.746} & 0.769 & \textbf{0.286} & \textbf{1} & \textbf{0.714} \\
\hline
Mistral & 17 (25) & 10 (10) & - & 0 (1) & 0 (2) & 0.557 & \textbf{0.87} & - & 0 & 0 \\
\hline
Wizardlm2 & 9 (51) & 1 (9) & - & 5 (6) & 3 (7) & 0.207 & 0.091 & - & 0.769 & 0.429 \\
\hline
Drain & 29 (44) & 9 (16) & 3 (80) & 6 (8) & 2 (4) & 0.725 & 0.621 & 0.07 & 0.8 & 0.364 \\
\hline
\end{tabular}
\label{results-performance}
\end{center}
\end{table*}

As for the number of detected templates for 10 experiment iterations, Drain always produced the same result since it is a deterministic algorithm. Also, by its design, Drain assigns all messages in the event log into groups, creating one-message groups for rare messages and thus leaving no messages unprocessed. 
{
As mentioned in the previous section, LLM-TD produced the same output during all experiment iterations, behaving similarly to Drain despite the non-deterministic nature of LLMs. 
}
{However}, the templates detected by LLM-TD might not cover the entire event log, and the number of uncovered messages was smallest for OpenChat (four messages for the \emph{sshd} data set and less for others{, see the row for OpenChat with \emph{k}=10 in Table \ref{results-runtime}}). 
In contrast, Wizardlm2 produced the largest number of uncovered messages for all data sets (e.g., in the case of \emph{sshd} and \emph{su} data sets {1121} and 822, respectively, which complicates their manual analysis).

{
As discussed in Section \ref{section_llmtd}, one notable difference of LLM-TD from other LLM-based template detection algorithms is its ability to identify multiple templates with one LLM query. This allows to reduce computationally expensive interaction with LLM, and we observed the effect of that property for OpenChat on three data sets. According to Tables \ref{results-llm-interaction} and \ref{results-performance}, OpenChat was able to detect 25 correct templates with 19 queries on the \emph{sshd} data set, 10 correct templates with 7 queries on the \emph{su} data set, and 7 correct templates with 3 queries on the \emph{snmpd} data set. Mistral featured a similar behavior on the \emph{su} data set. In contrast, because other LLM-based algorithms can identify at most one template with a single query, the detection of \emph{m} correct templates will require at least \emph{m} LLM queries, involving an additional computational overhead.
}

According to Table \ref{results-performance}, OpenChat achieved the best performance on four data sets, whereas Mistral had the highest F1-score on the \emph{su} data set. The \emph{suricata} data set was the most challenging for all evaluated approaches since Mistral and Wizardlm2 were not able to process it within 10 hours.
{
For \emph{k}=2 (the only setting where the processing of the \emph{suricata} data set completed in 10 hours), the F1-scores of Mistral and Wizardlm2 were 0.25 and 0, respectively.
}
{Furthermore,} OpenChat and Drain achieved low F1-scores of 0.286 and 0.07 {on the \emph{suricata} data set}. The vast majority of the log messages in this data set were Suricata IDS alerts with different numbers of words that correspond to one ground truth template (see Figure \ref{example-messages} in Section \ref{section_introduction}), and this complicated the detection of ground truth templates for all evaluated approaches. 
Another challenging data set was \emph{apache}, where only OpenChat achieved the F1-score of 0.714, whereas F1-scores of other approaches remained below 0.5. 

{
From the evaluated four methods, performance of Mistral and Wizardlm2 was noticeably lower. For example, Wizardlm2 exceeded the F1-score of 0.5 only in the case of the \emph{snmpd} data set. Also, Mistral yielded the F1-score of 0 on \emph{snmpd} and \emph{apache} data sets (the reason was the detection of overly general patterns, an issue which was already mentioned in the previous section and will be discussed in details in Section \ref{section_results_erroranalysis}).

Although \emph{k}=10 yielded the best F1-scores for the majority of experiment scenarios, we investigated if other values of \emph{k} would significantly improve the lower F1-scores from Table \ref{results-performance}. The F1-score of Mistral increased from 0 to 0.714 on the \emph{snmpd} data set with \emph{k}=5, whereas no improvement was observed for the \emph{apache} data set. As for Wizardlm2, its F1-score increased from 0.091 to 0.4 on the \emph{su} data set with \emph{k}=5, and from 0.207 to 0.357 on the \emph{sshd} data set with \emph{k}=20.
For other scenarios no performance improvements were observed.
}

As discussed in Section \ref{section_metrics}, we have used principle P1 to assess the correctness of detected templates. However, not employing this principle would allow us to assess the capabilities of different algorithms to infer ground truth templates from insufficient event log data. In Table \ref{results-decline}, relevant results are displayed.

\begin{table*}[htbp]
\caption{Performance of template detection methods without P1 (log message batch size 10)}
\begin{center}
\begin{tabular}{c c c c c c|c c c c c}
\hline
\textbf{Method} & \multicolumn{5}{c}{\textbf{Number of correctly detected templates}} & \multicolumn{5}{c}{\textbf{F1-score}} \\
\cline{2-11} 
& \textbf{\textit{sshd}} & \textbf{\textit{su}} & \textbf{\textit{suricata}} & \textbf{\textit{snmpd}} & \textbf{\textit{apache}} & \textbf{\textit{sshd}} & \textbf{\textit{su}} & \textbf{\textit{suricata}} & \textbf{\textit{snmpd}} & \textbf{\textit{apache}} \\
\hline
OpenChat & 17 & 10 & 3 & 5 & 3 & \textbf{0.507} & 0.769 & \textbf{0.286} & \textbf{0.714} & \textbf{0.429} \\
\hline
Mistral & 11 & 9 & - & 0 & 0 & 0.361 & \textbf{0.783} & - & 0 & 0 \\
\hline
Wizardlm2 & 7 & 1 & - & 4 & 0 & 0.161 & 0.091 & - & 0.615 & 0 \\
\hline
Drain & 9 & 5 & 0 & 4 & 0 & 0.225 & 0.345 & 0 & 0.533 & 0 \\
\hline
\end{tabular}
\label{results-decline}
\end{center}
\end{table*}

According to Table \ref{results-decline}, the performance of all methods deteriorated, and the decline in the F1-score was most notable for Drain. For example, for the \emph{sshd} data set, the F1-score decreased by 0.5, with the number of correctly detected templates decreasing from 29 to 9. Also, in the case of \emph{suricata} and \emph{apache} data sets, the number of correctly identified templates (and thus the F1-score) dropped to 0. Drain achieved its highest F1-score of 0.533 on the \emph{snmpd} data set.

For LLM-based methods, the decline in the performance of OpenChat was less severe, with its performance not deteriorating on \emph{su} and \emph{suricata} data sets. Also, OpenChat achieved the highest F1-score on {four} data sets. Mistral managed to reach the best {F1-score of 0.783 on the \emph{su} data set, with other F1-scores remaining below 0.4}. As for Wizardlm2, its F1-score dropped from 0.429 to 0 on the \emph{apache} data set{, whereas performance degraded less for other data sets.}

\begin{figure}[htbp]
\centerline{\includegraphics[width=0.5\textwidth]{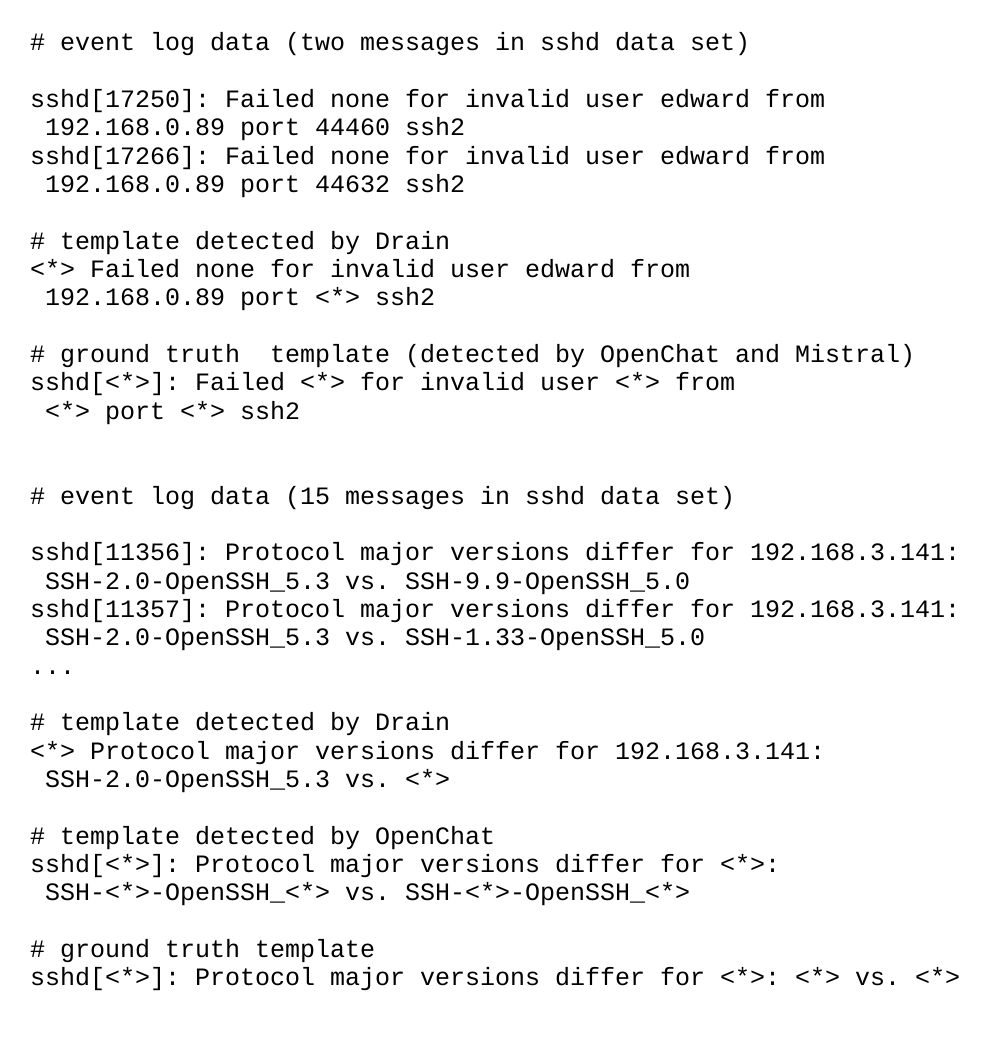}}
\caption{Inferring ground truth templates from insufficient log data}
\label{inferring-examples}
\end{figure}

These results illustrate that even if there is not enough information in log messages for inferring ground truth templates, some LLMs like OpenChat are able to achieve noticeably better results than traditional algorithms like Drain. To investigate that phenomenon, we analyzed detected templates and relevant log data manually, and Figure \ref{inferring-examples} displays two example cases to illustrate the following discussion.

In the case of the first scenario, OpenChat and Mistral were able to infer the ground truth template from insufficient event log data, whereas Drain produced a template that is not very informative (most variable fields are not properly identified). That example highlights one major advantage of LLM-based approaches over traditional algorithms, and we discovered a number of similar cases during our analysis, where templates detected by LLM-based approaches from insufficient log data were more informative. The superiority of LLM-based approaches over traditional algorithms comes from the fact that LLMs are trained on a very large set of widely different texts, allowing LLM-based methods to identify patterns that are not evident from the analyzed event log data.

The second scenario in Figure \ref{inferring-examples} illustrates the fact that although LLM-based approaches are not always able to produce a result that is identical to a ground truth template, the result can provide new insights into event log data by identifying regularities inside variable log message words. For example, the template by OpenChat in Figure \ref{inferring-examples} not only indicates that two SSH protocol fields are variables but also reveals that protocol strings of all messages begin with the prefix \emph{SSH-} and contain the substring \emph{-OpenSSH\_}. In contrast, traditional algorithms treat log message words as atoms and are not able to discover character patterns they might contain.

That advantage of LLM-based approaches over traditional algorithms is essential if ground truth templates have words that contain \wildcard{} as a substring, and the template mining algorithm is expected to identify them exactly as they appear in ground truth templates. For example, about 89\% of ground truth templates of data sets from Table \ref{datasets} have such a nature, and without using the P2 principle for assessing Drain's performance, the performance would decrease significantly. Also, without using neither P1 nor P2 principle, the performance of OpenChat, Mistral, and Wizardlm2 remained the same as reported in Table \ref{results-decline}, because we applied P2 only to Drain during the experiments. On the other hand, disabling both principles reduced the F1-score for Drain to 0 on all data sets, illustrating the strengths of LLM-based approaches over traditional algorithms when complex ground truth templates need to be identified.

{\subsubsection{Analysis of Incorrectly Detected Templates}\label{section_results_erroranalysis}}

As pointed out in a recent paper by Khan et al. \cite{khan}, existing works have not investigated incorrectly detected templates enough, and studying them helps to understand which types of detection errors are most common for specific algorithms. If $t$ is a template, let $log(t)$ denote the set of all potential messages (i.e., not only the messages in the evaluation data set!) which are matching template $t$. If $log(t_1) \subset log(t_2)$, then template $t_1$ is called \emph{more specific} than template $t_2$ (i.e., any message matched by $t_1$ is also matched by $t_2$, but not vice versa). Also, $t_2$ is called more general than $t_1$. According to \cite{khan}, an incorrect template is \emph{over-generalized} (OG) if it is more general than some ground truth template, and is \emph{under-generalized} (UG) if it is more specific than some ground truth template. If an incorrect template is neither OG nor UG, it is called \emph{mixed} (MX).

\begin{figure}[htbp]
\centerline{\includegraphics[width=0.5\textwidth]{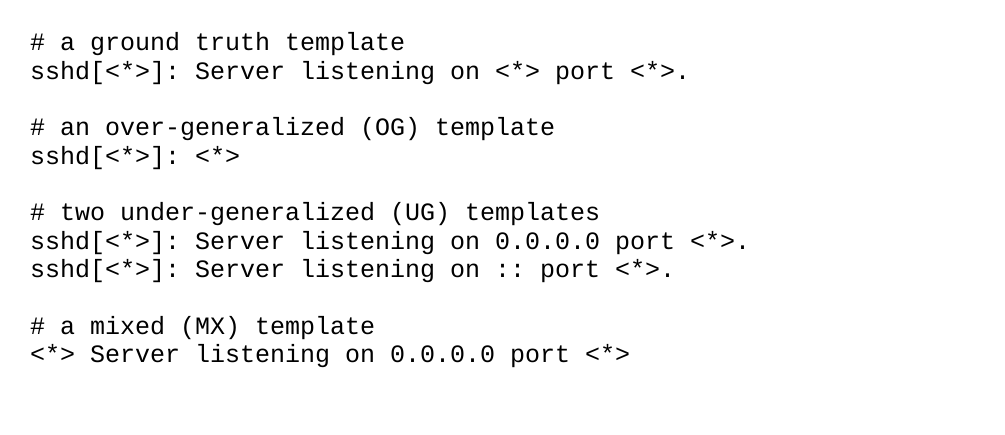}}
\caption{Types of incorrectly detected templates}
\label{og-ug-mx-examples}
\end{figure}

Figure \ref{og-ug-mx-examples} provides an example message for each type. For example, the last example from Figure \ref{og-ug-mx-examples} is of type MX since, unlike the ground truth template, it assumes the constant value 0.0.0.0 for the \emph{interface IP address} field. On the other hand, unlike the ground truth template, it does not assume that the message begins with the program name \emph{sshd} and ends with the dot character. Table \ref{incorrect-templates} provides data about the types of incorrect templates from our evaluation.

\begin{table*}[htbp]
\caption{Incorrectly detected templates by type (log message batch size 10)}
\begin{center}
\begin{tabular}{c c c c|c c c|c c c|c c c|c c c}
\hline
\textbf{Method} & \multicolumn{3}{c}{\textbf{sshd}} & \multicolumn{3}{c}{\textbf{su}} & \multicolumn{3}{c}{\textbf{suricata}} & \multicolumn{3}{c}{\textbf{snmpd}} & \multicolumn{3}{c}{\textbf{apache}} \\
\cline{2-16} 
& \textbf{\textit{OG}} & \textbf{\textit{UG}} & \textbf{\textit{MX}} & \textbf{\textit{OG}} & \textbf{\textit{UG}} & \textbf{\textit{MX}} & \textbf{\textit{OG}} & \textbf{\textit{UG}} & \textbf{\textit{MX}} & \textbf{\textit{OG}} & \textbf{\textit{UG}} & \textbf{\textit{MX}} & \textbf{\textit{OG}} & \textbf{\textit{UG}} & \textbf{\textit{MX}} \\
\hline
OpenChat & 2 & 4 & 0 & 2 & 1 & 0 & 0 & 12 & 0 & 0 & 0 & 0 & 2 & 0 & 0 \\
\hline
Mistral & 1 & 5 & 2 & 0 & 0 & 0 & - & - & - & 1 & 0 & 0 & 2 & 0 & 0 \\
\hline
Wizardlm2 & 0 & 42 & 0 & 0 & 8 & 0 & - & - & - & 0 & 0 & 1 & 0 & 4 & 0 \\
\hline
Drain & 1 & 1 & 13 & 1 & 2 & 4 & 0 & 43 & 34 & 0 & 2 & 0 & 2 & 0 & 0 \\
\hline
\end{tabular}
\label{incorrect-templates}
\end{center}
\end{table*}

As seen in Table \ref{incorrect-templates}, LLM-based approaches tend to produce very few MX templates, and for OpenChat no such templates were generated on five data sets. In contrast, for Drain, the MX templates formed a significant part among incorrectly detected templates on \emph{sshd}, \emph{su}, and \emph{suricata} data sets. As discussed above, traditional algorithms (including Drain) are not able to identify character patterns inside message words, and our investigation revealed that many incorrect templates from Drain represent the same scenario as illustrated by the MX template from Figure \ref{og-ug-mx-examples}.
{
However, on the \emph{snmpd} and \emph{apache} data sets, Drain produced just two incorrectly detected templates, and no prominent pattern of MX templates emerged.
}

When manually inspecting OG templates, we discovered that Mistral generated an OG template \emph{snmpd[\wildcard{}]: \wildcard{}} on the \emph{snmpd} data set during earlier phases of the event log processing. {Because} detected templates are used for filtering out log messages for performance optimization reasons (see line 4 in Algorithm \ref{llm-td} in Section \ref{section_llmtd}), no further messages were submitted to Mistral for analysis after the detection of this template{, although this would have allowed to find additional templates from the remaining event log data}. 
As 
{a result,} that led Mistral to identify only one template \emph{snmpd[\wildcard{}]: \wildcard{}}. 
{
On the \emph{apache} data set, Mistral suffered from the same issue, and the following two OG templates which matched all event log messages prevented the proper analysis of most event log data:

\emph{apache2: PHP Notice: \wildcard{} in \wildcard{} on line \wildcard{}}

\emph{apache2: PHP Warning: \wildcard{} in \wildcard{} on line \wildcard{}}

Also, as discussed in Section \ref{section_results_exectime}, the incorrect exclusion of log messages from processing by OG templates can have side effects on runtime, and allowed Mistral and Wizardlm2 to process the \emph{suricata} data set with the shortest execution times for \emph{k}=2.
}

Since other LLM-based template detection approaches such as LILAC \cite{lilac} and {Lunar} \cite{ulog} utilize similar work principles of employing already detected templates as filters, our finding illustrates one design limitation of current LLM-based approaches. Namely, if an underlying LLM is prone to produce overly general templates, a mechanism must be implemented to identify and drop such templates. Unfortunately, whereas some overly general templates like the OG template from Figure \ref{og-ug-mx-examples} are fairly easy to detect, there are other more subtle cases, and their identification is a non-trivial task. 
{
For example, aforementioned two OG templates detected by Mistral on the \emph{apache} data set contain six words and three \wildcard{}, giving them a look of a regular template and hiding their too general nature.
}

\begin{figure}[htbp]
\centerline{\includegraphics[width=0.5\textwidth]{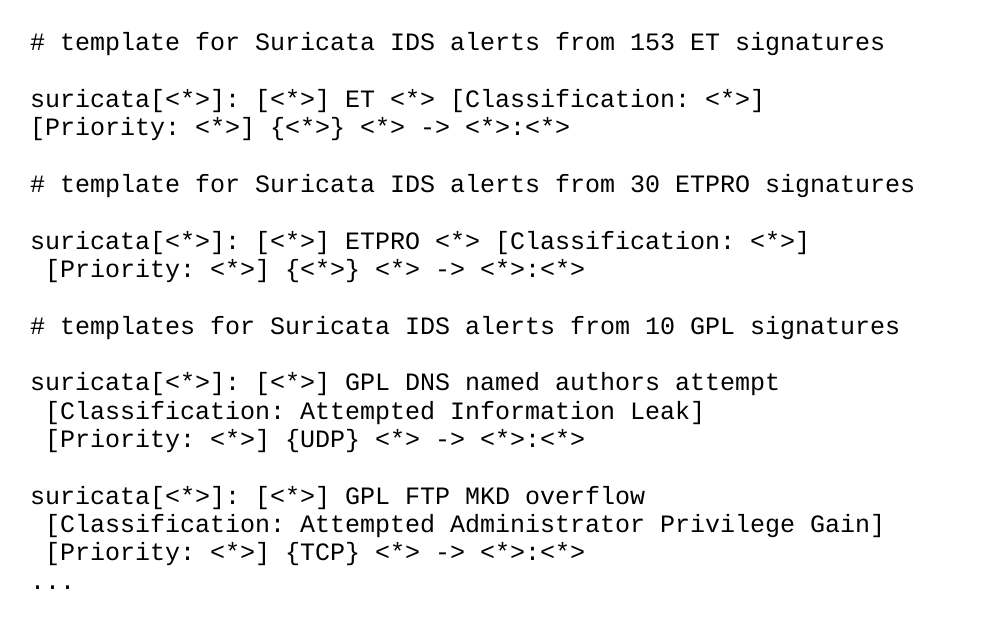}}
\caption{Templates from OpenChat describing Suricata IDS alerts}
\label{suricata-templates}
\end{figure}

When investigating UG templates, we found that some of them provide novel insights into the event log data despite being incorrect according to ground truth.
The most notable case involved the \emph{suricata} data set, which was the most challenging for all algorithms because Mistral and Wizardlm2 failed to process it, whereas the highest F1-score remained below 0.3 and was achieved by OpenChat (see Table \ref{results-performance}). In the case of OpenChat, all incorrectly identified templates were of type UG and have been shown in Figure \ref{suricata-templates}. The total number of UG templates was 12, and they were more specific than the ground truth template shown in Figure \ref{example-messages} (Section \ref{section_introduction}), which covered over 99\% of the \emph{suricata} data set.

Also, in Suricata IDS alert messages (see Figure \ref{example-messages} in Section \ref{section_introduction}), the first word of the alert string identifies the signature supplier. For example, in two alerts from Figure \ref{example-messages} the supplier identifiers are \emph{ET} (free signatures from EmergingThreats) and \emph{GPL} (community signatures). Another identifier present in the \emph{suricata} data set is \emph{ETPRO} (commercial signatures from EmergingThreats).

Despite being incorrectly identified, the UG templates from Figure \ref{suricata-templates} reveal the fact that IDS alerts have been produced by signatures from three suppliers. Since for \emph{ET} and \emph{ETPRO}, a larger number of signatures produced many different alert strings, alerts for both suppliers have been summarized by a single template. However, since for the \emph{GPL} supplier, only 10 signatures triggered alerts, individual templates have been created for each \emph{GPL} signature, providing more details for the human analyst. Since OpenChat did not generate any other incorrectly identified templates, templates from Figure \ref{suricata-templates} can be regarded as a concise and meaningful summary of Suricata IDS alerts, providing useful insights not present in ground truth. 

As for Drain (the other algorithm that managed to process the \emph{suricata} data set), it neither detected the ground truth template for IDS alerts nor produced a meaningful summary for them. When investigating 77 incorrectly detected templates from Drain, we found that the issue was related to the common design limitation of traditional algorithms, which assume that the log messages matching the same template contain the same number of words. Since Drain has this particular design limitation, it could not detect meaningful templates from Suricata IDS alert data.

\begin{figure}[htbp]
\centerline{\includegraphics[width=0.5\textwidth]{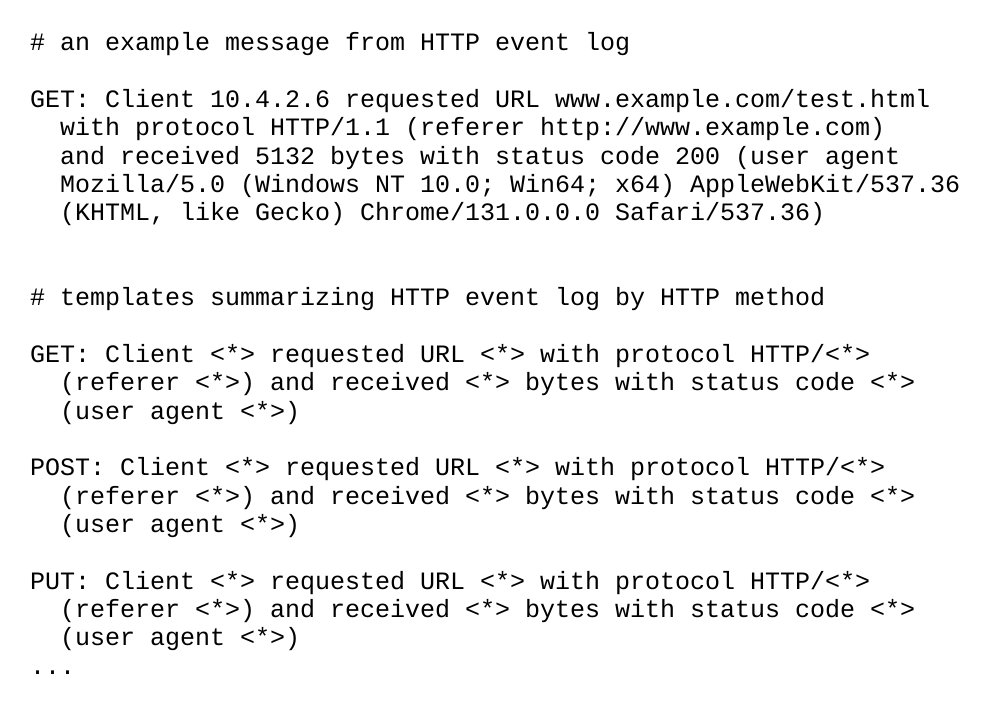}}
\caption{Templates from OpenChat describing HTTP requests}
\label{http-templates}
\end{figure}

{
Finally, to estimate the capabilities of LLM-TD to process large non-\emph{syslog} data sets, we applied it to a textual event log of all HTTP requests observed in a large organizational network. The event log contained 15,311,709 messages describing HTTP requests from 36,821 clients, and all messages were following the same format. Figure \ref{http-templates} displays an example anonymized message from this event log (since the log data had a confidential nature, we were not able to release this data set publicly).
}

{
When applying LLM-TD with OpenChat to this event log 10 times, it was processed within 6--7 minutes despite its large size of 3.9 GB, producing identical output for all 10 iterations. Furthermore, the memory consumption of LLM-TD remained modest, ranging from 65.5 to 67.4 MB across all iterations (we observed similar memory footprints lower than 70 MB for all other experiments described earlier in this paper). 

Although OpenChat was not able to detect a single template for all messages, it was able to meaningfully summarize them with 11 templates (see Figure \ref{http-templates}), which represented distinct HTTP methods identified in the event log (for one HTTP method, two templates were created which reflected different usage patterns of this method). However, 17 log messages remained uncovered by templates from Figure \ref{http-templates} and were reported to the end user. When inspecting them more closely, we discovered that six messages represented previously unknown malicious HTTP requests with binary bytes used in the HTTP method field.
}

Scenarios from Figure \ref{suricata-templates}{, Figure \ref{http-templates},} and Figure \ref{inferring-examples} demonstrate that LLM-based approaches like OpenChat have a potential not only for template detection but also for \emph{discovering previously unknown knowledge} that is not represented by ground truth. Also, it highlights the fact that detected templates deserve qualitative analysis during evaluations to better assess the knowledge discovery capabilities of different algorithms.

\section{Discussion}\label{section_discussion}

In this section, we will summarize the main findings of the evaluation presented in Section \ref{section_evaluation}.

First, as discussed in Section \ref{section_metrics}, existing performance metrics are not entirely suited for a meaningful comparison of LLM-based and traditional algorithms since the latter have several design limitations. Also, adjusting the ground truth templates according to these limitations (e.g., as proposed in \cite{khan}) would not allow us to evaluate the unique capabilities of LLM-based approaches. As a solution, we have proposed two heuristic principles for assessing the correctness of detected templates.

Second, our results show that local LLMs with parameter size 7B can feature a good performance when used without fine-tuning, employing the in-context learning paradigm in an unsupervised fashion. In particular, the LLM-TD algorithm with the OpenChat LLM featured the best performance among all evaluated methods, including Drain, a highly capable traditional algorithm \cite{zhu}. That allows security analysts to employ local LLMs for unsupervised event log analysis without the need to create labeled data sets. Moreover, since security event logs contain highly sensitive data, they can be analyzed without submitting sensitive information to external service providers.

Whereas existing LLM-based algorithms \cite{lilac, ulog, divlog, logppt, llmparser} have attempted to detect one template during each LLM query, the LLM-TD algorithm presented in this paper utilizes LLMs for extracting several templates at once during every query. As the results from Section \ref{section_results} indicate, such a detection strategy is a viable alternative to existing approaches{, and it allows to reduce computationally expensive interaction with LLMs.}

One drawback of LLM-based template detection algorithms is their need for more computational resources than traditional algorithms require, which leads to longer execution times. For example, during our evaluations, a high-end GPU was used by local LLMs (see Section \ref{section_experiment_env}), whereas traditional algorithms do not need such dedicated hardware. In addition, Drain featured significantly faster execution times than LLM-TD with OpenChat, Mistral, and Wizardlm2 (see Table \ref{results-runtime} in Section \ref{section_results_exectime}). Also, in the case of one challenging data set (\emph{suricata}), Mistral and Wizardlm2 were not able to process it even within 10 hours {for most log message batch sizes}.

Furthermore, LLM-based approaches have several unique capabilities that traditional algorithms lack. According to our findings (e.g., see Figure \ref{inferring-examples} in Section \ref{section_results_tempdetection}), LLMs can infer informative templates even from insufficient event log data. Also, LLMs are able to detect character patterns inside log message words which can yield novel insights into event log data. 

Our study has also identified several design limitations of LLM-based and traditional algorithms, which can interfere with the template detection process. According to recent research literature \cite{divlog, gasimov}, many traditional algorithms incorrectly assume that log messages matching the same template have the same number of words, and our study has confirmed the negative impact of this assumption. Also, since traditional algorithms consider the log message words as atoms, detected templates are often less informative. As for the limitations of the LLM-based algorithms, they commonly employ already detected templates to exclude log messages from further analysis. However, if overly general templates are identified during the early phases of event log processing, a large fraction of relevant messages could be left unprocessed.

Finally, as our study has indicated, qualitative analysis of detected templates is important since templates that deviate from ground truth can provide novel insights into event log data. Unfortunately, existing studies have not paid enough attention to this aspect. However, as our study has shown, LLM-based algorithms are able to find previously unknown knowledge from event log data, and this warrants the qualitative analysis of \emph{all} detected templates.

\section{Conclusion and Future Work}\label{section_future_work}

In the current paper, we have presented the LLM-TD algorithm for unsupervised template detection from security event logs, and have demonstrated that small local LLMs can be successfully used for template discovery purposes. Furthermore, our study has pointed out the drawbacks of existing performance metrics, and has proposed several heuristic principles for addressing these drawbacks. The current paper has also highlighted the advantages of LLM-based template detection over traditional algorithms, and has emphasized the importance of qualitative analysis of all identified templates. The following paragraphs will outline some research directions for future work. 

In previous studies, a generic wildcard notation (e.g., \wildcard{}) has been used for denoting a variable part in log messages that match the template, and the current paper has adopted the same approach. However, such variable parts can have a different nature. For example, in Figure \ref{example-messages} (Section \ref{section_introduction}), \wildcard{} can match IP addresses, port and process numbers, user names, etc. Instead of using a generic \wildcard{} notation, more specific notations for different data types (e.g., $\langle*:number\rangle$ or $\langle*:ipaddress\rangle$) would make detected templates significantly more informative. Whereas traditional algorithms would need regular expression-based event log preprocessing to achieve that goal (e.g., see \cite{logcluster, drain}), LLMs have the potential of detecting such templates from the original event log data, and studying approaches of finding these templates is a valuable further research direction.

In addition to detecting more informative templates, LLMs could be utilized to discover regular expressions that correspond to these templates since human analysts often employ them instead of templates for actual event log processing tasks. However, according to our preliminary experiments, discovering regular expressions directly from event log messages using LLMs is a complex task. To address that issue, LLMs could be tasked with detecting some simplified regular-expression-like intermediate formats from log messages so that these intermediate formats could be converted into regular expressions with a little effort.

{
One of the limitations of the current study is the template detection performance analysis for a limited number of local LLMs. Since new LLMs are constantly released, evaluating the template detection capabilities of a larger number of local LLMs is an open future research topic.
}

Finally, as mentioned in Section \ref{section_introduction}, there is a shortage of publicly available recent security event logs, which would facilitate the cyber security-centric event log analysis research. Although this paper has published several such event logs, creating additional realistic security event log data sets and releasing them into the public domain remains an important task for future work.

\begin{acknowledgements}
This work was supported by the Estonian Centre of Excellence in Artificial Intelligence (EXAI), funded by the Estonian Ministry of Education and Research grant TK213.
\end{acknowledgements}

\end{document}